\documentclass[english,12pt,aps,prd,a4paper,preprintnumbers,floatfix,nofootinbib,showpacs,superscriptaddress, notitlepage]{revtex4-1} 
\usepackage[mode=buildnew]{standalone}
\usepackage[usenames,dvipsnames]{color}  
\usepackage{graphicx}

\usepackage{setspace}
\usepackage{caption}
\usepackage{subcaption}
\usepackage{amsmath}
\usepackage{amssymb}
\usepackage[colorlinks=true,citecolor=darkred,urlcolor=darkred, pdfborder={0 0 0}]{hyperref}
\usepackage[normalem]{ulem}
\usepackage{xcolor}
\usepackage{placeins}
\usepackage{mathrsfs}
\usepackage{verbatim} 
\usepackage{cancel} 
\usepackage{float} 

\usepackage{scalerel}
\usepackage{tikz-feynman}
\tikzfeynmanset{compat=1.1.0}
\usepackage{feynmp}
\usepackage{tikzsymbols}
\usepackage{multirow}

\makeatletter
\def\p@subsection{}
\makeatother

\definecolor{darkred}{rgb}{0.6,0,0}

\definecolor{linkcolor}{rgb}{0,0,0.5}

\def\gsim{\raise0.3ex\hbox{$\;>$\kern-0.75em\raise-1.1ex\hbox{$\sim\;$}}}
\def\lsim{\raise0.3ex\hbox{$\;<$\kern-0.75em\raise-1.1ex\hbox{$\sim\;$}}}

\def\beqn#1{\begin{equation}\label{#1}}
\def\eeqn{\end{equation}}

\def\beqa#1{\begin{eqnarray}\label{#1}}
\def\eeqa{\end{eqnarray}}

\newcommand {\ignore}[1]{}

\def\Z4{$Z_4$}
\def\O5{$\mathcal{O}_5$ }

\def\321{$\mathrm{SU(3) \otimes SU(2) \otimes U(1)}$ }

\usepackage{booktabs}
\usepackage{slashed}

\begin{document}

\title{\color{Blue} From LUX-ZEPLIN to Colliders: Probing Higgsino Dark Matter
}

\author{Kingman Cheung}\email{cheung@phys.nthu.edu.tw}
\affiliation{Department of Physics, National Tsing Hua University, Hsinchu 30013, Taiwan}
\affiliation{Center for Theory and Computation, National Tsing Hua University, Hsinchu 30013, Taiwan}

\author{Sin Kyu Kang}
\email{skkang@seoultech.ac.kr}
\affiliation{Seoul National University of Science and Technology, Seoul 01811, Republic of Korea}
 \author{Ranjeet Kumar}\email{kumarranjeet.drk@gmail.com}
 \affiliation{Institute for Convergence of Basic Studies, Seoul National University of Science and Technology, Seoul 01811, Republic of Korea}

\begin{abstract}
The high-energy nuclear recoil event with recoil energy $E_R \approx 248\text{ keV}$ observed 
by the LZ collaboration provides an exciting hint toward a 
model with a $1.1\text{ TeV}$ Higgsino inelastic dark matter.
Such a recoil energy requires a mass splitting of order 
($\delta \approx 350\text{ keV}$) between the two nearly-degenerate
neutral states.
We show that within the framework of the MSSM, the model predicts 
a nearly-degenerate charged Higgsino state, chargino, whose mass 
splitting from the neutral states is of order $\mathcal{O}(350)$~MeV.
The subsequent decays of such charginos once produced at colliders 
would lead to a sub-centimeter charged track or tracklet 
(with lifetime $\tau \approx 0.025\text{ ns}$) at the detector.
We show that such a scenario is not constrained by the current LHC bounds, and could be tested at future high-energy colliders, including
HL-LHC, 100 TeV $pp$ colliders, and muon colliders. 
The most critical requirement is how short a track or tracklet can
be reconstructed.
\end{abstract}

\maketitle


\section{Introduction} \label{sec:intro}

A broad range of astrophysical and cosmological observations points to the existence of dark matter (DM) \cite{Zwicky:1933gu,Rubin:1970zza,Rubin:1980zd,Planck:2018vyg}, but its particle nature and cosmic origin remain elusive.
Among the various mechanisms proposed for generating the observed DM abundance, thermal freeze-out has been studied extensively~\cite{Scherrer:1985zt}. In this framework, DM particles are initially in thermal equilibrium with the Standard Model (SM) plasma, but eventually decouple as the Universe expands and the interaction rate falls below the Hubble rate, leaving behind a relic abundance. The most prominent realization of this mechanism involves weakly interacting massive particles (WIMPs)~\cite{Lee:1977ua,Kolb:1990vq,Jungman:1995df}.
WIMPs can span a broad mass range, extending from the keV scale up to roughly $100$ TeV. 
The simplicity of the WIMP paradigm, together with its weak but potentially observable interactions with ordinary matter, makes it particularly attractive for direct detection searches.

Direct detection experiments provide an avenue for probing such interactions. In this regard, the recent LUX-ZEPLIN (LZ) analysis~\cite{LZ2026HighRecoil} extends the nuclear recoil search window to approximately $270~\mathrm{keV}$, providing new sensitivity to DM interactions with recoil spectra that are suppressed at low energies and enhanced in the high recoil regime.
Using an exposure of $2.84$ tonne-years and a nuclear recoil search window extending up to approximately $270~\mathrm{keV}$, the LZ analysis reported a single event with characteristics at
\begin{equation} \label{eq:LZevent}
  E_R = 248\pm 23\,(\mathrm{stat})\pm 23\,(\mathrm{sys})~\mathrm{keV}.
\end{equation}
The collaboration examined a range of potential backgrounds and detector related effects but found no plausible explanation for such an observed event.
  A profile-likelihood analysis gives a global background-only significance of $2.6\sigma$ after accounting for the look-elsewhere effect, with a maximum local significance of $3.4\sigma$ over the interaction models tested
\cite{LZ2026HighRecoil}.  While the observation falls well short of the statistical significance required for a discovery, the unusually high recoil energy of the single event makes it an interesting case for investigating possible particle physics explanations. Such interpretations should, however, be regarded as exploratory given the limited event statistics.

Among the possible interpretations, endothermic inelastic DM is particularly well suited to the high recoil regime, where the inelastic mass splitting leads to distinctive scattering kinematics.
In endothermic inelastic scattering, a DM particle $\chi_1$ in the Galactic halo transitions to a slightly heavier state $\chi_2$, with $\delta=m_{\chi_2}-m_{\chi_1}>0$. The minimum incident velocity required to produce a nuclear recoil of energy $E_R$ from a target nucleus of mass $m_N$ is
\begin{equation}
 v_{\min}(E_R)=
 \frac{1}{\sqrt{2m_NE_R}}
 \left(
 \frac{m_NE_R}{\mu_{\chi N}}+\delta
 \right),
 \label{eq:vmin}
\end{equation}
where $\mu_{\chi N}$ is the DM-nucleus reduced mass.  A positive splitting therefore suppresses
low-energy scattering and pushes the observable spectrum toward the high-energy tail.  Generic
analyses of the LZ event find representative masses above several hundred GeV and splittings of
order a few hundred keV \cite{SuYangYang2026}.
Interestingly, the LZ analysis itself considers an inelastic spin-independent interaction resembling a Higgsino model as one of its illustrative scenarios \cite{Graham:2024syw}. This motivates a closer examination of the Higgsino framework as a concrete realization of the required inelastic spectrum.

A nearly pure Higgsino provides a particularly economical realization. In the heavy gaugino limit, the two neutral Higgsinos form a pseudo-Dirac pair that is split into two Majorana states.  Their coupling to the $Z$ boson is predominantly off-diagonal, so the leading tree-level $Z$ interaction mediates
\begin{equation}
 \widetilde{\chi}_1^0 + N \rightarrow \widetilde{\chi}_2^0 + N.
 \label{eq:inelastic}
\end{equation}
Recent studies have shown that a nearly pure Higgsino with a mass close to the thermal value, $m_{\widetilde \chi^0_{1,2}} \sim1~\mathrm{TeV}$, and a neutral state splitting of order a few hundred keV can provide an intriguing interpretation of the LZ high recoil event~\cite{Freese:2026sga,Su:2026rwz,Yin:2026jnn,Du:2026guj}.\footnote{See also Refs.~\cite{Wu:2026nhi,Fan:2026kxx,Lou:2026idn,Visinelli:2026kgt,Yamashita:2026ump,DiMauro:2026ldr,Nomura:2026qyq,Smirnov:2026aqk,Unwin:2026rdp,McCabe:2026crm,Jeesun:2026vzo,Rodd:2026tyn,Chattopadhyay:2026ryw,Dent:2026bji,Gu:2026vto,deLima:2026shq} for other recent interpretations of the LZ high recoil event in terms of inelastic DM and related scenarios.}
An important consequence of this interpretation is that the nearly-degenerate neutral Higgsinos necessarily 
implies a nearly-degenerate charged state, the lightest chargino $\widetilde{\chi}_1^\pm$, as they all 
originate from the same $\mu$ parameter. 
While the neutral state splitting relevant for LZ is of order hundreds of keV, 
the charged-neutral splitting is typically of order a few hundred MeV due mainly to electroweak radiative corrections. 
The resulting chargino predominantly decays into a neutral Higgsino and a soft charged pion, and 
the sub-leading decays are into a neutral Higgsino and a charged lepton plus neutrino. 
Since the mass splitting is so small that the chargino could lead to a charged track before decaying, 
a characteristic short disappearing-track or a tracklet 
signature at colliders.

In this work, we explore this collider implication of the LZ-motivated Higgsino scenario. We study the relation between the neutral Higgsino splitting required by inelastic direct detection and the charged Higgsino spectrum that determines the chargino lifetime, and examine the resulting prospects for short-track searches at present and future colliders. This provides a complementary test of the Higgsino interpretation of the LZ event, since direct detection probes the neutral state splitting whereas collider experiments are sensitive to the charged-neutral mass difference and the associated chargino decay length.

\section{The Higgsino Dark Matter Framework}
\label{sec:Higgsino_framework}

We now turn to the Higgsino DM framework, which provides a simple and well-motivated realization of the inelastic DM scenario of interest. We consider the nearly pure Higgsino limit of the Minimal Supersymmetric Standard Model (MSSM), where the relevant neutral and charged states exhibit the characteristic mass structure governing their phenomenology.

\subsection{Thermal Higgsino DM and inelastic scattering}

A nearly pure Higgsino constitutes one of the simplest electroweak DM candidates in the MSSM. In the limit in which the bino and wino are significantly heavier than the Higgsino mass parameter,
\begin{equation}
|M_1|,\ |M_2| \gg |\mu|,
\end{equation}
the lightest electroweakino states consist of two nearly degenerate neutral Higgsinos,
$\widetilde\chi_1^0$ and $\widetilde\chi_2^0$, together with a charged Higgsino,
$\widetilde\chi_1^\pm$.
For standard thermal freeze-out, Higgsino annihilation and co-annihilation through electroweak interactions determine the relic abundance. Requiring the Higgsino to account for the observed DM abundance,
$\Omega_{\rm DM}h^2\simeq0.12$ \cite{Planck:2018vyg}, 
yields a mass close to
\begin{equation}
m_{\widetilde \chi^0_{!,2} }\simeq |\mu|\simeq (1.0-1.1)~{\rm TeV},
\label{eq:thermal_Higgsino_mass}
\end{equation}
up to the usual dependence on radiative corrections and the rest of the supersymmetric spectrum. We adopt $m_{\widetilde \chi^0_1 }\simeq1.1~{\rm TeV}$ as the representative thermal benchmark in the following.
Nonthermal cosmological histories can allow lighter Higgsinos without requiring them to obtain the observed relic abundance through standard freeze-out. Such scenarios are phenomenologically interesting, particularly in light of the recently discussed high-energy-sideband constraint, but we restrict the present analysis to the predictive thermal benchmark $m_{\tilde \chi^0_1 }\simeq1.1~{\rm TeV}$.

In the exact Higgsino number limit, the two neutral Higgsino Weyl fields combine into a Dirac fermion. Mixing with the heavy electroweak gauginos after electroweak symmetry breaking breaks this degeneracy and produces two Majorana mass eigenstates. The corresponding $Z$ interaction is predominantly off-diagonal in the neutral Higgsino mass basis,
\begin{equation}
{\cal L}_Z
\supset
\frac{g}{2c_W}
Z_{\mu}
\overline{\widetilde\chi_2^0}\gamma^\mu
\widetilde\chi_1^0
+\mathrm{h.c.},
\label{eq:offdiagonal_Z}
\end{equation}
up to convention dependent phases in the Majorana fields. The important physical point is that the unsuppressed tree-level $Z$ current connects the two different neutral Higgsino states, while the corresponding diagonal vector current vanishes in the pure Higgsino limit.
The leading nuclear scattering process is therefore naturally inelastic,
\begin{equation}
\widetilde\chi_1^0+N
\rightarrow
\widetilde\chi_2^0+N.
\label{eq:inelastic_Higgsino}
\end{equation}

We define the neutral Higgsino mass splitting as
\begin{equation}
\delta m_0
\equiv
m_{\widetilde\chi_2^0}
-
m_{\widetilde\chi_1^0}>0.
\end{equation}
The minimum velocity for Higgsino inelastic scattering is then given by Eq.~(\ref{eq:vmin}) with the identification $\delta=\delta m_0$. The positive splitting introduces an endothermic threshold, suppressing scattering from the bulk of the Galactic velocity distribution and preferentially selecting its high-velocity tail. As a result, the recoil spectrum can extend to substantially higher energies than in conventional elastic WIMP scattering.
For a nearly pure thermal Higgsino with a mass around $\sim 1$ TeV as provided in Eq.~(\ref{eq:thermal_Higgsino_mass}), analyses of the LZ event indicate that the predicted inelastic scattering rate approaches the region of interest for a neutral state splitting of order a few hundred keV, with a representative value
\begin{equation}
\delta m_0
\sim 350~{\rm keV}.
\label{eq:LZ_delta0}
\end{equation}
The precise preferred value depends on the assumed Galactic velocity distribution and experimental acceptance, but the essential feature is the sub-MeV neutral state splitting.

We note that a recent analysis has pointed out a potential tension with the Higgsino interpretation arising from the higher-energy $S1c$ sideband of the LZ data. A thermal Higgsino capable of producing the observed $248~{\rm keV}$ recoil may also generate events at still higher recoil energies, whereas no events were observed in the corresponding sideband. At present, however, the experimental acceptance in this region has not been publicly provided, preventing a definitive exclusion of the Higgsino interpretation. We therefore regard the LZ event as a motivated benchmark rather than evidence for Higgsino DM, while keeping this potential sideband constraint in mind.


\subsection{Origin of the neutral Higgsino splitting} \label{subsec:split}
The neutral Higgsino mass splitting originates from the mixing of the Higgsino states with the heavier gauginos.
The neutralino mass matrix in the basis
$(\widetilde B,\widetilde W^0,\widetilde H_d^0,\widetilde H_u^0)^T$ is given as follows
\begin{equation}
{\cal M}_N=
\begin{pmatrix}
M_1 & 0 & -m_Zs_Wc_\beta & m_Zs_Ws_\beta \\
0 & M_2 & m_Zc_Wc_\beta & -m_Zc_Ws_\beta \\
-m_Zs_Wc_\beta & m_Zc_Wc_\beta & 0 & -\mu \\
m_Zs_Ws_\beta & -m_Zc_Ws_\beta & -\mu & 0
\end{pmatrix}.
\label{eq:neutralino_mass_matrix}
\end{equation}
In the heavy gaugino limit, integrating out the bino and wino generates the leading Higgsino-number-violating operators. At leading order in $m_Z/M_{1,2}$, the neutral splitting is approximately given by
\begin{equation}
\delta m_0
\simeq
m_Z^2
\left(
\frac{s_W^2}{M_1}
+
\frac{c_W^2}{M_2}
\right)
+
{\cal O}
\left(
\frac{\mu m_Z^2}{M_{1,2}^2}
\right).
\label{eq:neutral_mass_split}
\end{equation}

Equation~(\ref{eq:neutral_mass_split}) shows that reproducing the LZ-motivated splitting in Eq.~(\ref{eq:LZ_delta0}) is highly constraining. In the absence of cancellations between the bino and wino contributions, the required gaugino scale is parametrically
$M_{1,2}\sim{\cal O}(10^7~{\rm GeV})$,
far above the Higgsino mass scale. 
Alternatively, if $M_1$ and $M_2$ have opposite signs, the bino and wino contributions can partially cancel, allowing a sub-MeV neutral splitting to be realized with intermediate-scale gaugino masses. In this cancellation region, however, the leading order expression in Eq.~(\ref{eq:neutral_mass_split}) is no longer sufficiently accurate for determining a splitting at the few hundred keV level.
Indeed, near the cancellation line, the residual leading order contribution becomes comparable to the nominally subleading terms of order $\mu m_Z^2/M_{1,2}^2$. Thus, Eq.~(\ref{eq:neutral_mass_split}) remains useful for illustrating the parametric origin of the cancellation, but the physical neutral Higgsino splitting must be obtained from the full neutralino mass matrix.
We therefore numerically diagonalize the mass matrix in Eq.~(\ref{eq:neutralino_mass_matrix}). 
The condition $\delta m_0\simeq350~{\rm keV}$ admits two solutions, $M_2\simeq128.28~{\rm TeV}$ and $M_2\simeq130.10~{\rm TeV}$, on the two sides of the cancellation point at $M_2\simeq129.18~{\rm TeV}$. In the following, we adopt the upper solution and define the representative intermediate-scale benchmark as \cite{Du:2026guj}
\begin{equation}
\mu=1.091~{\rm TeV},\qquad
M_1=-38.6~{\rm TeV},\qquad
M_2=130.10~{\rm TeV},\qquad
\tan\beta=10.
\label{eq:intermediate_benchmark}
\end{equation}
This benchmark yields $\delta m_0\simeq350~{\rm keV}$.
This distinction will be relevant below because the neutral Higgsino splitting and the charged-neutral splitting depend on different combinations of the underlying electroweakino parameters.


\section{Chargino Mass Splitting and Decay}
\label{sec:chargino_properties}

The Higgsino spectrum also contains a charged state $\widetilde\chi_1^\pm$, whose mass and decay properties are important for collider phenomenology. We therefore examine the charged-neutral mass splitting and the resulting chargino decay in this section.

\subsection{Charged-neutral Higgsino splitting}

The charged Higgsino mass is determined by the chargino mass matrix,
\begin{equation}
{\cal M}_C=
\begin{pmatrix}
M_2 & \sqrt{2}m_W s_\beta \\
\sqrt{2}m_W c_\beta & \mu
\end{pmatrix}.
\label{eq:chargino_matrix}
\end{equation}
For $|M_2|\gg|\mu|$, the lightest chargino is dominantly Higgsino and is nearly degenerate with the two neutral Higgsinos at tree level.
This degeneracy is lifted by electroweak radiative corrections. In the asymptotic pure Higgsino limit, the one-loop electroweak contribution approaches
\begin{equation}
\Delta m_+^{\rm EW}
\equiv
m_{\widetilde\chi_1^\pm}
-
m_{\widetilde\chi_1^0}
\simeq
(340-350)~{\rm MeV},
\label{eq:chargino_loop_split}
\end{equation}
with a commonly quoted value around $350~{\rm MeV}$ for a TeV scale Higgsino. More precise calculations including finite electroweak masses and higher order effects give a value in this range.
It is important that the charged-neutral splitting is parametrically much larger than the neutral state splitting selected by the LZ interpretation. Comparing Eqs.~(\ref{eq:LZ_delta0}) and (\ref{eq:chargino_loop_split}), one finds
$\delta m_0 \ll \Delta m_+$,
with a hierarchy of roughly three orders of magnitude.

The charged splitting need not coincide exactly with the pure Higgsino radiative value when the gaugino masses are finite. For the representative intermediate-scale benchmark defined in Eq.~(\ref{eq:intermediate_benchmark}), an exact tree-level diagonalization of the neutralino and chargino mass matrices gives
\begin{equation}
\Delta m_+^{\rm tree}\simeq-9.2~{\rm MeV}.
\label{eq:massplit}
\end{equation}
This correction is small compared with, but not entirely negligible relative to, the electroweak radiative contribution of approximately $(340-350)~{\rm MeV}$ for a TeV scale Higgsino. At the level of the present approximation, one therefore expects the physical charged-neutral splitting to lie in the vicinity of
$
\Delta m_+\sim(330-340)~{\rm MeV}.
$
We emphasize that this estimate should not be interpreted as a precision pole-mass prediction. A consistent calculation of the physical charged-neutral mass difference requires the complete one-loop result, including finite gaugino contributions and the associated renormalization-scheme dependence. For the collider considerations below, the robust implication is a charged Higgsino splitting of order $300~{\rm MeV}$, with the corresponding chargino lifetime discussed below.
%
\subsection{Chargino decay and lifetime}

The charged-neutral mass splitting determines the dominant decay modes and lifetime of the chargino. In our scenario, the mass ordering is
\begin{equation}
m_{\widetilde\chi_1^\pm}
>
m_{\widetilde\chi_2^0}
>
m_{\widetilde\chi_1^0},
\end{equation}
allowing the chargino to decay into either of the two neutral Higgsino states.
Since
$\delta m_0\sim{\cal O}(10^{-4})~{\rm GeV}$ is negligible compared with the charged splitting,
the two decay modes have essentially identical kinematics at collider resolution. We therefore denote the neutral Higgsino in the final state generically by $\widetilde\chi^0$.
For $\Delta m_+>m_{\pi^\pm}$,
the dominant decay proceeds through an off-shell $W$ boson into a single charged pion,
\begin{equation}
\widetilde\chi_1^\pm
\rightarrow
\widetilde\chi^0 \; \pi^\pm,
\label{eq:chargino_pion_decay}
\end{equation}
while the leptonic three-body channels
\begin{equation}
\widetilde\chi_1^\pm
\rightarrow
\widetilde\chi^0 \; \ell^\pm\nu_\ell,
\qquad
\ell=e,\mu,
\label{eq:chargino_leptonic}
\end{equation}
provide smaller contributions.

In the heavy Higgsino limit, the dominant one-pion contribution has the characteristic dependence
\begin{equation}
\Gamma_\pi
\simeq
\frac{G_F^2|V_{ud}|^2f_\pi^2}{\pi}
(\Delta m_+)^3
\sqrt{
1-\frac{m_{\pi^\pm}^2}{(\Delta m_+)^2}
},
\label{eq:chargino_width}
\end{equation}
where the normalization corresponds to the summed Higgsino neutral state contribution in the nearly degenerate limit. Here we use $f_\pi\simeq130.2~{\rm MeV}$.
Precision lifetime predictions require the inclusion of the subleading leptonic modes, hadronic corrections, and electroweak radiative effects.
For
$\Delta m_+\simeq (340-350)~{\rm MeV}$,
the proper lifetime lies approximately in the range
\begin{equation}
\tau_{\widetilde\chi_1^\pm}
\sim
(2-3)\times10^{-11}~{\rm s}
\simeq
(0.02-0.03)~{\rm ns},
\label{eq:chargino_lifetime}
\end{equation}
corresponding to
\begin{equation}
c\tau_{\widetilde\chi_1^\pm}
\sim
(6-8)~{\rm mm}.
\label{eq:chargino_ctau}
\end{equation}
The charged pion produced in Eq.~(\ref{eq:chargino_pion_decay}) is also very soft. Neglecting corrections of order $\Delta m_+/m_{\widetilde \chi^0}$, its momentum in the chargino rest frame is
\begin{equation}
p_\pi^\ast
\simeq
\sqrt{
(\Delta m_+)^2
-
m_{\pi^\pm}^2
}.
\label{eq:pion_momentum}
\end{equation}
For the LZ-motivated Higgsino spectrum, we get
\begin{equation}
p_\pi^\ast
\simeq
(300-320)~{\rm MeV}.
\label{eq:pion_momentum_numerical}
\end{equation}
The pion is therefore typically too soft to provide a conventional high-$p_T$ object, whereas the chargino can travel a macroscopic distance of several millimeters before decaying.

The resulting spectrum can be summarized schematically as
\begin{align}
m_{\widetilde\chi_1^0}
&\simeq1.1~{\rm TeV},
\quad 
m_{\widetilde\chi_2^0}
-
m_{\widetilde\chi_1^0}
\sim350~{\rm keV},
\nonumber \\ 
m_{\widetilde\chi_1^\pm}
-
m_{\widetilde\chi_1^0}
&\sim (340-350)~{\rm MeV},
\quad
c\tau_{\widetilde\chi_1^\pm}
\sim (6-8)~{\rm mm}.
\label{eq:spectrum_summary}
\end{align}
This combination of a TeV scale electroweak state, a sub-centimeter chargino decay length, and an ${\cal O}(300~{\rm MeV})$ visible pion defines the collider target implied by the LZ-motivated Higgsino interpretation. 
The spectrum summarized in Eq.~(\ref{eq:spectrum_summary}) defines the collider target examined in the next section.

\section{Collider Searches and Future Prospects}
\label{sec:collider}

The LZ-motivated Higgsino spectrum summarized in Eq.~(\ref{eq:spectrum_summary}) presents a particularly challenging collider target. The charged Higgsino is sufficiently long-lived to travel a macroscopic distance, but typically too short-lived to generate a conventional reconstructed charged particle track. At the same time, its electroweak production cross section is strongly suppressed by the TeV scale mass. The combination of the short decay length and the small production rate explains why the thermal Higgsino region motivated by the LZ event remains well beyond the current LHC sensitivity.

\subsection{Production and collider signature}
\label{subsec:production}

At hadron colliders, Higgsino-like electroweakinos are produced predominantly through Drell-Yan processes,
\begin{align}
pp \rightarrow
\widetilde\chi_1^+
\widetilde\chi_1^-,
\qquad
pp \rightarrow
\widetilde\chi_1^\pm
\widetilde\chi_{1,2}^0,
\qquad
pp \rightarrow
\widetilde\chi_1^0
\widetilde\chi_2^0 .
\label{eq:Higgsino_production}
\end{align}
For a common Higgsino mass of approximately $1.1~{\rm TeV}$, the inclusive electroweakino production cross section at $\sqrt{s}=13~{\rm TeV}$ is only
\begin{equation}
\sigma_{\rm incl}
\left(
pp\rightarrow\widetilde H\widetilde H
\right)
\simeq 0.57~{\rm fb},
\label{eq:Higgsino_xsec_13}
\end{equation}
at NLO+NNLL accuracy~\cite{LHCSUSYXSWG}.
Here $\tilde{H}$ stands for a neutral or charged Higgsino state.
Thus, even before detector acceptance and analysis requirements, only ${\cal O}(10^2)$ Higgsino events are produced in the full Run-2 data set.

Because the visible products from the compressed Higgsino decays are very soft, the Higgsino system itself does not normally provide an efficient trigger. Searches therefore rely on an energetic initial-state-radiation (ISR) object recoiling against the electroweakino pair, leading to the characteristic topology
\begin{equation}
pp\rightarrow
\widetilde H\widetilde H+j_{\rm ISR}
\longrightarrow
j_{\rm ISR}
+\slashed E_T
+{\rm short~tracklet}
+\pi_{\rm soft}^{\pm}.
\label{eq:collider_topology}
\end{equation}
The hard ISR jet serves two purposes. It generates the missing transverse momentum trigger and, at the same time, boosts the Higgsino system, thereby increasing the distance traveled by the chargino in the laboratory.
For a chargino with transverse momentum $p_T$, the average transverse decay length is
\begin{equation}
\left\langle L_T\right\rangle
=
\frac{p_T}{m_{\widetilde\chi_1^\pm}}
c\tau_{\widetilde\chi_1^\pm}.
\label{eq:transverse_length}
\end{equation}
Correspondingly, the probability for the chargino to survive beyond a transverse radius $R$ is
\begin{equation}
P(L_T>R)=
\exp\left[
-\frac{R m_{\widetilde\chi_1^\pm}}
{p_Tc\tau_{\widetilde\chi_1^\pm}}
\right].
\label{eq:survival_probability}
\end{equation}
Equation~(\ref{eq:survival_probability}) is particularly useful for understanding the experimental difficulty of the LZ-motivated benchmark.

\subsection{Present disappearing track sensitivity}
\label{subsec:current_LHC}

The most relevant LHC constraint comes from searches for disappearing or very short charged tracks \cite{Fukuda:2017jmk}. The latest ATLAS analysis uses $137~{\rm fb}^{-1}$ of Run-2 data at $\sqrt{s}=13~{\rm TeV}$ and reconstructs exceptionally short tracks from only three or four measurements in the innermost tracking layers~\cite{ATLAS2026Disappearing}. For the three-hit category, ATLAS further separates events according to whether a low-energy pion associated with the chargino decay is reconstructed using a dedicated algorithm.
The short proper lifetime in Eq.~(\ref{eq:spectrum_summary}) nevertheless causes a severe geometric suppression. To illustrate this, the four barrel pixel layers of the Run-2 ATLAS detector are located approximately at
\begin{equation}
R_{\rm pixel}
\simeq
33.3~{\rm mm},\ 50.5~{\rm mm},\ 88.5~{\rm mm},\ 122.5~{\rm mm}.
\label{eq:pixel_radii}
\end{equation}
Thus, in the central region, a track reconstructed from three pixel measurements typically requires the chargino to survive to a radius of order
$R_{3{\rm hit}}\sim 9~{\rm cm}$.

\begin{figure}[!]
\centering
\includegraphics[width=\textwidth]{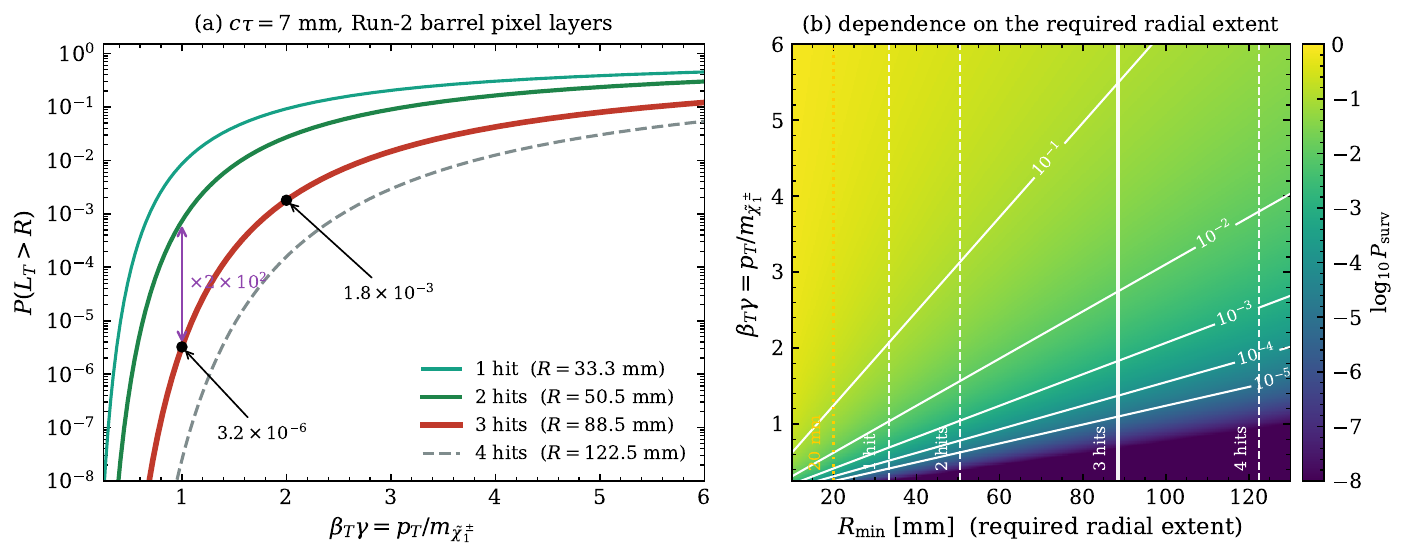}
\caption{Geometric acceptance for a chargino with $c\tau = 7$~mm.
\textbf{(a)} Survival probability of Eq.~(\ref{eq:survival_probability}) versus the transverse boost
$\beta_T\gamma$. Each curve is labeled by the number
of Run-2 barrel pixel hits obtained, with the three-hit requirement in red.
\textbf{(b)} $\log_{10}P_{\rm surv}$ as a function of the required radial extent
$R_{\min}$ and the boost, with iso-probability contours; vertical lines mark one
to four pixel hits and, dotted, a $20$~mm vertex detector radius. Geometric
acceptance only: the production spectrum, efficiencies and fake-track background
are not included.}
\label{fig:survival}
\end{figure}

The geometric suppression implied by Eq.~(\ref{eq:survival_probability}) is displayed in
Fig.~\ref{fig:survival}.
Panel (a) shows the survival probability as a function of
the transverse boost
\begin{equation}
\beta_T\gamma \equiv \frac{p_T}{m_{\tilde\chi^\pm_1}} ,
\end{equation}
evaluated at the LZ-motivated proper decay length $c\tau = 7$~mm. The relevant
quantity is not the position of any single detector layer but the radial extent a
chargino must reach in order to satisfy the hit multiplicity required by the
analysis: with the Run-2 barrel radii of Eq.~(\ref{eq:pixel_radii}), one, two, three and four pixel
hits demand survival past $33.3$, $50.5$, $88.5$ and $122.5$~mm respectively. For
the three-hit requirement the probability is $3.2\times10^{-6}$ at
$\beta_T\gamma = 1$ and reaches only $1.8\times10^{-3}$ at $\beta_T\gamma = 2$. For
$m_{\tilde\chi^\pm_1} = 1.1$~TeV the latter already corresponds to
$p_T \simeq 2.2$~TeV, so that reconstructible three-hit tracklets originate almost
entirely from the extreme boosted tail of the production spectrum.

Panel (b) generalizes this to an arbitrary required radial extent $R_{\min}$.
Because the exponent in Eq.~(\ref{eq:survival_probability}) is linear in $R_{\min}$ and inversely linear in
$\beta_T\gamma$, the iso-probability contours are straight lines through the
origin,
\begin{equation}
\beta_T\gamma = \frac{R_{\min}}{n\ln(10) c\tau}
\qquad\text{for}\qquad P_{\rm surv} = 10^{-n},
\label{eq:isoP}
\end{equation}
so that the geometric acceptance depends only on the ratio $R_{\min}/c\tau$. At
fixed boost, relaxing the requirement from three pixel hits ($R_{\min} = 88.5$~mm)
to two ($R_{\min} = 50.5$~mm) enhances the survival probability by
\begin{equation}
\frac{P(50.5~{\rm mm})}{P(88.5~{\rm mm})}
= \exp\left[\frac{38.0~{\rm mm}}{\beta_T\gamma~ c\tau}\right] ,
\end{equation}
which is a factor of $2\times10^{2}$ at $\beta_T\gamma = 1$. This illustrates why
improvements in short-track reconstruction and in tracker geometry can provide
gains that are difficult to achieve through increased luminosity alone.

Two caveats should be attached to this statement. First, the gain quoted above is
itself boost dependent, falling to a factor of $15$ at $\beta_T\gamma = 2$ and to
$6$ at $\beta_T\gamma = 3$: the benefit of reducing $R_{\min}$ is largest precisely
in the low-boost region where the production rate is largest but the ISR-based
trigger is least efficient, and it shrinks in the boosted tail that current
analyses actually select. Second, Fig.~\ref{fig:survival} quantifies the geometric
acceptance only. A complete sensitivity estimate must fold in $d\sigma/dp_T$,
the trigger and reconstruction efficiencies, and the fake-tracklet background,
which grows as $R_{\min}$ is reduced. The competition between these effects is
precisely what a dedicated study of the benchmark in Eq.~(\ref{eq:intermediate_benchmark}) would have to
resolve.
This observation is more informative than simply comparing $c\tau$ with the radius of the first pixel layer. Although the chargino has a macroscopic proper lifetime, the probability of surviving through enough detector layers to form a reconstructible tracklet is exponentially small. The ISR requirement helps by preferentially selecting boosted events, but at the price of further reducing the already small signal cross section.

The latest ATLAS disappearing track analysis excludes pure Higgsino charginos up to approximately
$m_{\widetilde\chi_1^\pm}\simeq225~{\rm GeV}$
for lifetimes below about $0.03~{\rm ns}$~\cite{ATLAS2026Disappearing}. This lifetime range directly overlaps with the prediction of the LZ-motivated Higgsino scenario, but the corresponding mass reach remains far below the thermal value of $1.1~{\rm TeV}$.
An independent and complementary strategy has recently been pursued by CMS using a low momentum isolated track together with large missing transverse momentum~\cite{CMS2026SoftTrack}. This search is specifically sensitive to the soft charged pion produced in compressed electroweakino decays. For Higgsino-like spectra with
$0.28~{\rm GeV}
\lesssim\Delta m_+
\lesssim1.15~{\rm GeV}$,
CMS excludes chargino masses up to approximately $185~{\rm GeV}$. The LZ-motivated value $\Delta m_+\sim340~{\rm MeV}$ lies directly within the targeted splitting range, but again the $1.1~{\rm TeV}$ thermal mass is well outside the present sensitivity.
The large gap between the present LHC sensitivity and the LZ-motivated thermal Higgsino benchmark is summarized in Table~\ref{tab:current_searches}.
\begin{table}[!h]
\centering
\begin{tabular}{|l|c|c|c|}
\hline
Search &
Data set &
Target regime &
Approximate Higgsino reach\\
\hline 
ATLAS disappearing track
&
$137~{\rm fb}^{-1}$
&
$\tau_{\chi^\pm}\lesssim0.03~{\rm ns}$
&
$m_{\chi^\pm}\lesssim225~{\rm GeV}$
\\
CMS soft isolated track
&
$138~{\rm fb}^{-1}$
&
$\Delta m_+=(0.28-1.15)~{\rm GeV}$
&
$m_{\chi^\pm}\lesssim185~{\rm GeV}$
\\
LZ-motivated Higgsino
&
-
&
$c\tau\simeq(6-8)~{\rm mm}$
&
$m_{\chi^\pm}\simeq1.1~{\rm TeV}$
\\
\hline
\end{tabular}
\caption{Representative current LHC searches relevant to the LZ-motivated Higgsino benchmark. The quoted limits refer to the corresponding simplified Higgsino scenarios and should not be interpreted as model independent mass bounds.}
\label{tab:current_searches}
\end{table}
Heavy stable charged particle searches are not directly relevant to this benchmark. Such analyses are optimized for particles whose lifetimes are sufficiently long to traverse a substantial fraction, or all, of the tracking and muon systems. The Higgsino considered here instead decays predominantly within the innermost tracking volume. Thus, the experimentally relevant regime lies between conventional prompt compressed spectrum searches and searches for detector stable charged particles.

\subsection{Prospects at the HL-LHC}
\label{subsec:HLLHC}

The High-Luminosity LHC (HL-LHC) offers an important opportunity to extend the collider sensitivity to the LZ-motivated Higgsino scenario, with an anticipated integrated luminosity of approximately
$3000~{\rm fb}^{-1}$
at $\sqrt{s}=14~{\rm TeV}$. However, the LZ-motivated thermal Higgsino remains a difficult target because increasing luminosity alone does not remove the exponentially small track survival probability in Eq.~(\ref{eq:survival_probability}).
This observation suggests that the most important improvements are not simply higher luminosity, but rather an optimization of the analysis and detector response specifically for sub-centimeter lifetimes. This points to several directions in which the sensitivity could be improved.
First, reducing the minimum radial distance required for a reconstructible track can produce a parametrically large increase in signal acceptance. As follows directly from Eq.~(\ref{eq:survival_probability}), the survival probability depends exponentially on the ratio $R_{\rm min}/(\beta_T\gamma c\tau)$. Even a modest reduction in $R_{\rm min}$ can therefore substantially increase the acceptance for a $c\tau\sim7~{\rm mm}$ chargino. Two-hit or hit-level tracklet strategies therefore deserve particular attention for this benchmark.

Second, the terminating chargino trajectory can be correlated with the soft pion from the decay $\widetilde\chi_1^\pm\rightarrow\widetilde\chi^0\pi^\pm$, whose characteristic momentum in the chargino rest frame is $p_\pi^\ast\sim300~{\rm MeV}$, as shown in the previous section (see Eq.~(\ref{eq:pion_momentum_numerical})). Although this makes conventional reconstruction difficult, the pion can exhibit a nonzero transverse impact parameter associated with the displaced chargino decay. Combining a very short tracklet with a geometrically correlated displaced soft track can therefore provide significant additional background rejection. The latest ATLAS three-hit plus soft-pion strategy already represents an important step in this direction~\cite{ATLAS2026Disappearing}.

Third, the ISR requirement can be optimized not only as a trigger condition but also as a lifetime selection variable. Increasing the transverse recoil boosts the chargino system and increases the survival probability. An optimized analysis should therefore balance the rapidly decreasing hard-ISR production rate against the exponentially increasing tracklet efficiency.
These considerations imply that a dedicated HL-LHC study of the LZ-motivated benchmark in Eq.~(\ref{eq:spectrum_summary}) would be valuable. Existing HL-LHC disappearing-track projections were generally developed before the most recent three-hit and soft-pion reconstruction techniques. A quantitative extrapolation incorporating the upgraded inner tracker, realistic fake-track backgrounds, and the new short-track algorithms is therefore required before assigning a robust HL-LHC mass reach to the LZ-motivated scenario.

\subsection{A 100 TeV proton collider}
\label{subsec:FCC}

A future proton collider with $\sqrt{s}=100~{\rm TeV}$ would substantially increase both the Higgsino production rate and the available boost. The latter is particularly important for a short-lived Higgsino, since the collider sensitivity is controlled not only by the total production cross section but also by the fraction of charginos that reach the required tracking radius.
A detailed disappearing track study for a future $100~{\rm TeV}$ collider found that the thermal Higgsino region can be probed with an integrated luminosity of order
\begin{equation}
{\cal L}_{\rm int}\sim30~{\rm ab}^{-1},
\end{equation}
provided that the inner tracking system and short-track reconstruction are appropriately optimized~\cite{SaitoEtAl2019}. In particular, that study considered a pure Higgsino with
\begin{equation}
m_{\widetilde H}\simeq1~{\rm TeV},
\qquad
\tau_{\chi^\pm}\simeq0.023~{\rm ns},
\end{equation}
which closely matches the lifetime region of interest here, and found that a $5\sigma$ discovery of a $1~{\rm TeV}$ Higgsino is feasible under favorable tracker configurations.

These studies also highlight the importance of detector geometry for probing this regime. Moving the relevant inner tracking layers closer to the interaction point can enhance the signal acceptance by orders of magnitude for the short-lived Higgsino case. Pixel-level timing information at the ${\cal O}(50~{\rm ps})$ level can additionally suppress fake-track backgrounds in the high-pileup environment. Thus, a $100~{\rm TeV}$ collider appears capable of directly testing the thermal Higgsino interpretation, but this conclusion relies on designing the detector with short disappearing tracks explicitly in mind.

\subsection{Prospects at a high-energy muon collider}
\label{subsec:muon_collider}

A high-energy muon collider offers a complementary environment for the same Higgsino spectrum. Chargino pairs can be directly produced through
\begin{equation}
\mu^+\mu^-
\rightarrow
\widetilde\chi_1^+
\widetilde\chi_1^-,
\label{eq:muc_production}
\end{equation}
primarily through $s$-channel electroweak interactions. In contrast to a proton collider, the partonic center-of-mass energy is fixed, and the production kinematics can therefore be considerably cleaner.
A detailed detector level study of disappearing tracks at a $10~{\rm TeV}$ muon collider with
\begin{equation}
{\cal L}_{\rm int}=10~{\rm ab}^{-1}
\end{equation}
found that a thermal Higgsino with
\begin{equation}
m_{\widetilde H}\simeq1.1~{\rm TeV},
\qquad
\tau_{\chi^\pm}\simeq0.02~{\rm ns},
\end{equation}
can be probed at the $5\sigma$ level~\cite{CapdevillaEtAl2021}. The dominant experimental challenge is the large beam-induced background generated by muon decays. However, timing information and spatial correlations between neighboring detector hits were shown to suppress fake tracklets to a manageable level.

The short Higgsino lifetime makes the radial position of the innermost tracking layers particularly important also at a muon collider. Interestingly, the minimal reconstructible tracklet length considered in Ref.~\cite{CapdevillaEtAl2021} lies close to the proper decay scale of the pure Higgsino. Consequently, the LZ-motivated spectrum provides a concrete benchmark for future vertex detector design.
We stress, however, that the statement that a $10~{\rm TeV}$ muon collider can discover the thermal Higgsino is conditional on the assumed luminosity, detector layout, and control of beam-induced backgrounds. It should therefore be interpreted as a demonstrated projection rather than a guaranteed experimental outcome.

\subsection{Complementarity of present and future searches}
\label{subsec:collider_summary}

The collider phenomenology follows directly from the Higgsino spectrum motivated by the LZ high recoil event. A neutral state splitting of $\delta m_0\sim{\cal O}(350~{\rm keV})$, together with a thermal Higgsino mass near $1.1~{\rm TeV}$, leads to the charged state properties summarized in Eq.~(\ref{eq:spectrum_summary}). At hadron colliders, the resulting compressed Higgsino system is searched for using a hard initial-state-radiation jet, giving the characteristic topology introduced in Eq.~(\ref{eq:collider_topology}).
\begin{figure}[t]
\centering
\includegraphics[width=0.86\textwidth]{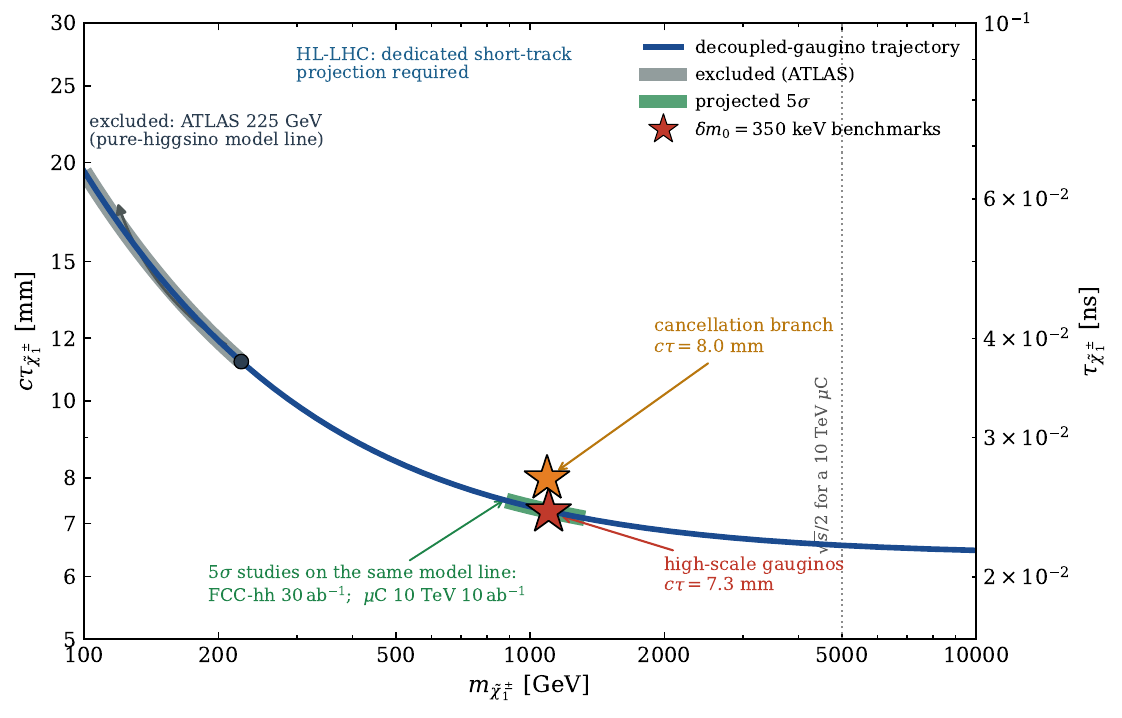}
\caption{Chargino decay length versus mass. The blue curve is the pure Higgsino
trajectory in the decoupled-gaugino limit. Stars mark the
two realisations of $\delta m_0 \simeq 350$~keV of Sec.~\ref{subsec:split}: high-scale gauginos,
lying on the curve, and the cancellation branch, displaced upward by $\Delta m_+^{\rm tree} \simeq -9$~MeV. Grey and green bands indicate the portions of the trajectory excluded by ATLAS and covered at $5\sigma$ by the $100$~TeV and
$10$~TeV muon collider studies; these are reaches quoted for the same model line,
not digitized two-dimensional contours. The dotted line is $\sqrt{s}/2$ for a
$10$~TeV muon collider.}
\label{fig:mctau}
\end{figure}

At the present LHC, the combination of a sub-fb production cross section and the very small probability of surviving through several pixel layers leaves the $1.1~{\rm TeV}$ benchmark far beyond existing sensitivity. The HL-LHC will provide a much larger event sample, but reaching this benchmark will depend critically on improvements in short-track reconstruction rather than luminosity alone. A $100~{\rm TeV}$ hadron collider and a multi-TeV muon collider, on the other hand, have been shown in dedicated studies to have sensitivity to the thermal Higgsino mass scale.

The LZ-motivated scenario is therefore particularly useful as a benchmark connecting direct detection and collider detector design. Unlike generic long-lived particle searches, in which the mass and lifetime are commonly scanned independently, the Higgsino interpretation predicts a rather narrow target around $m_{\widetilde\chi_1^\pm}\sim1.1~{\rm TeV}$ and $c\tau_{\widetilde\chi_1^\pm}\sim{\cal O}(7~{\rm mm})$. A collider search optimized specifically for this region would provide a direct and astrophysics independent test of the particle physics interpretation of the LZ high recoil event.
It is worth emphasizing how strongly constrained this target is. 
In the decoupled-gaugino pure Higgsino limit, the chargino mass and lifetime are not independent: the charged-neutral splitting is fixed by the electroweak loop as shown in 
Eq.~(\ref{eq:chargino_loop_split}), and the decay length then 
follows from Eq.~(\ref{eq:chargino_width}). The
prediction is therefore a one-parameter trajectory in the
$(m_{\tilde\chi^\pm_1}, c\tau)$ plane rather than a region,
\begin{equation}
  c\tau_{\tilde\chi^\pm_1}
  \;=\; c\tau\!\left[\Delta m_+\!\left(m_{\tilde\chi^\pm_1}\right)\right],
  \label{eq:line}
\end{equation}
shown as the solid curve in Fig.~\ref{fig:mctau}. It is remarkably flat, varying
only between about $20$~mm at $m_{\tilde\chi^\pm_1} = 100$~GeV and $6.5$~mm at
$10$~TeV, because $\Delta m_+$ approaches its asymptotic value already at the TeV
scale. Finite gaugino effects perturb this trajectory through
Eq.~(\ref{eq:massplit}), so the LZ-motivated benchmark need not lie exactly on it: the two
realizations of $\delta m_0 \simeq 350$~keV discussed in Sec.~\ref{subsec:split} are shown
separately, and the intermediate-scale cancellation benchmark sits about $10\%$
above the decoupled gaugino curve. Finite gaugino corrections can therefore shift
the chargino lifetime away from the pure Higgsino trajectory, potentially
providing information complementary to the neutral state splitting measured in
direct detection.

The trajectory also organizes the collider information. Because the searches
relevant here quote their reach for the pure Higgsino model, in which $c\tau$ is
tied to the mass by Eq.~(\ref{eq:line}), their results are statements about
particular points of this curve rather than about an independent
$(m_{\tilde\chi^\pm_1}, c\tau)$ region. The ATLAS limit of $225$~GeV excludes the
low-mass portion of the trajectory, indicated in grey; the $100$~TeV and $10$~TeV
muon collider studies establish $5\sigma$ sensitivity in the region around
$1$~TeV, indicated in green, which contains the LZ-motivated benchmarks. We stress
that these are model line statements: the underlying sensitivities are
two-dimensional contours in the mass-lifetime plane, and a quantitative
comparison would require digitizing them from each analysis. For the same reason,
we do not display an HL-LHC region, since, as noted in Sec.~\ref{subsec:HLLHC}, a robust projection
for this benchmark does not yet exist. The CMS low-momentum track search is omitted from the figure because its quoted reach is obtained at a mass splitting
away from the pure Higgsino trajectory and cannot be placed on it. The corresponding result is instead included in Table~\ref{tab:current_searches}.

The same thermal Higgsino benchmark is also being probed through indirect detection. In particular, electroweak annihilation of a $1.1~{\rm TeV}$ Higgsino produces gamma-ray line and endpoint signatures in the TeV energy range. Recent H.E.S.S. searches have begun to probe the thermal Higgsino parameter space for sufficiently cuspy Galactic DM profiles, while CTAO is expected to provide substantially improved sensitivity. These probes are complementary to the collider searches considered here, although their interpretation necessarily depends on the Galactic DM density profile. We do not pursue the indirect detection phenomenology further in this work.

\section{Conclusion}
The high-energy nuclear recoil at $E_R \approx 248\text{ keV}$ observed by LZ provides an exciting hint toward $1.1\text{ TeV}$ Higgsino inelastic DM. 
In the MSSM realization considered here, mixing with the heavy bino and wino generates a neutral state mass splitting already at tree level, with $\delta m_0 \simeq 350~\mathrm{keV}$ taken as a representative benchmark.
The charged Higgsino is separated from the lightest neutral state by a much larger mass difference of order a few hundred MeV, dominated by electroweak radiative corrections for the benchmarks considered here, with an additional tree-level contribution from
finite-gaugino mixing.
The subsequent decay of such charginos at colliders would lead to
a sub-centimeter charged track or tracklet (with lifetime $\tau \approx 0.025\text{ ns}$). 
While completely evading current LHC bounds, this scenario serves as a concrete benchmark for 
testing this Higgsino DM model 
at future high-energy colliders and 
next-generation direct detection experiments.


\section*{Acknowledgments} 
\noindent
The work of KC is supported by the National Science \& Technology Council under grant no. NSTC 113-2112-M-007-041-MY3.  SKK and RK are supported by the National Research Foundation of Korea under Grant NRF-2023R1A2C100609111. We acknowledge the use of Claude for assistance with figure generation and visualization during the preparation of this work.

\bibliographystyle{utphys}
\bibliography{references.bib} 

\providecommand{\href}[2]{#2}\begingroup\raggedright\begin{thebibliography}{10}

\bibitem{Zwicky:1933gu}
F.~Zwicky, ``{Die Rotverschiebung von extragalaktischen Nebeln},''
  \href{http://dx.doi.org/10.1007/s10714-008-0707-4}{{\em Helv. Phys. Acta}
  {\bfseries 6} (1933) 110--127}.

\bibitem{Rubin:1970zza}
V.~C. Rubin and W.~K. Ford, Jr., ``{Rotation of the Andromeda Nebula from a
  Spectroscopic Survey of Emission Regions},''
  \href{http://dx.doi.org/10.1086/150317}{{\em Astrophys. J.} {\bfseries 159}
  (1970) 379--403}.

\bibitem{Rubin:1980zd}
V.~C. Rubin, N.~Thonnard, and W.~K. Ford, Jr., ``{Rotational properties of 21
  SC galaxies with a large range of luminosities and radii, from NGC 4605 /R =
  4kpc/ to UGC 2885 /R = 122 kpc/},''
  \href{http://dx.doi.org/10.1086/158003}{{\em Astrophys. J.} {\bfseries 238}
  (1980) 471}.

\bibitem{Planck:2018vyg}
{\bfseries Planck} Collaboration, N.~Aghanim {\em et~al.}, ``{Planck 2018
  results. VI. Cosmological parameters},''
  \href{http://dx.doi.org/10.1051/0004-6361/201833910}{{\em Astron. Astrophys.}
  {\bfseries 641} (2020) A6}, \href{http://arxiv.org/abs/1807.06209}{{\ttfamily
  arXiv:1807.06209 [astro-ph.CO]}}. [Erratum: Astron.Astrophys. 652, C4
  (2021)].

\bibitem{Scherrer:1985zt}
R.~J. Scherrer and M.~S. Turner, ``{On the Relic, Cosmic Abundance of Stable
  Weakly Interacting Massive Particles},''
  \href{http://dx.doi.org/10.1103/PhysRevD.33.1585}{{\em Phys. Rev. D}
  {\bfseries 33} (1986) 1585}. [Erratum: Phys.Rev.D 34, 3263 (1986)].

\bibitem{Lee:1977ua}
B.~W. Lee and S.~Weinberg, ``{Cosmological Lower Bound on Heavy Neutrino
  Masses},'' \href{http://dx.doi.org/10.1103/PhysRevLett.39.165}{{\em Phys.
  Rev. Lett.} {\bfseries 39} (1977) 165--168}.

\bibitem{Kolb:1990vq}
E.~W. Kolb and M.~S. Turner,
  \href{http://dx.doi.org/10.1201/9780429492860}{{\em {The Early Universe}}},
  vol.~69.
\newblock Taylor and Francis, 5, 2019.

\bibitem{Jungman:1995df}
G.~Jungman, M.~Kamionkowski, and K.~Griest, ``{Supersymmetric dark matter},''
  \href{http://dx.doi.org/10.1016/0370-1573(95)00058-5}{{\em Phys. Rept.}
  {\bfseries 267} (1996) 195--373},
  \href{http://arxiv.org/abs/hep-ph/9506380}{{\ttfamily arXiv:hep-ph/9506380}}.

\bibitem{LZ2026HighRecoil}
{LZ Collaboration}, ``Search for dark matter particle interactions in an
  extended nuclear recoil energy window with the lux-zeplin (lz) experiment,''
  \href{http://arxiv.org/abs/2609.02823}{{\ttfamily arXiv:2609.02823
  [hep-ex]}}.

\bibitem{SuYangYang2026}
L.~Su, J.~M. Yang, and W.-N. Yang, ``Inelastic dark matter signature at high
  recoil energy in lux-zeplin and cresst,''
  \href{http://arxiv.org/abs/2609.01475}{{\ttfamily arXiv:2609.01475
  [hep-ph]}}.

\bibitem{Graham:2024syw}
P.~W. Graham, H.~Ramani, and S.~S.~Y. Wong, ``{Enhancing direct detection of
  Higgsino dark matter},''
  \href{http://dx.doi.org/10.1103/PhysRevD.111.055030}{{\em Phys. Rev. D}
  {\bfseries 111} no.~5, (2025) 055030},
  \href{http://arxiv.org/abs/2409.07768}{{\ttfamily arXiv:2409.07768
  [hep-ph]}}.

\bibitem{Freese:2026sga}
K.~Freese and D.~P. Theodosopoulos, ``{Higgsino Dark Matter Interpretation of
  the LUX-ZEPLIN 248 keV Nuclear-Recoil Event},''
  \href{http://arxiv.org/abs/2609.01583}{{\ttfamily arXiv:2609.01583
  [hep-ph]}}.

\bibitem{Su:2026rwz}
L.~Su, J.~M. Yang, and W.-N. Yang, ``{Inelastic Dark Matter Signature at High
  Recoil Energy in LUX-ZEPLIN and CRESST},''
  \href{http://arxiv.org/abs/2609.01475}{{\ttfamily arXiv:2609.01475
  [hep-ph]}}.

\bibitem{Yin:2026jnn}
W.~Yin, ``{A PQ-Symmetric High-Scale SUSY Interpretation of the LZ High-Energy
  Recoil},'' \href{http://arxiv.org/abs/2609.01892}{{\ttfamily arXiv:2609.01892
  [hep-ph]}}.

\bibitem{Du:2026guj}
X.~Du and F.~Wang, ``{TeV Higgsino Interpretation of the LZ High-Recoil Event
  with Intermediate-Scale Electroweak Gauginos},''
  \href{http://arxiv.org/abs/2609.04163}{{\ttfamily arXiv:2609.04163
  [hep-ph]}}.

\bibitem{Wu:2026nhi}
L.~Wu, Y.~Zhang, and B.~Zhu, ``{TeV Higgsino Dark Matter from LZ Nuclear Recoil
  to Fermi-LAT Gamma Rays},'' \href{http://arxiv.org/abs/2609.01590}{{\ttfamily
  arXiv:2609.01590 [hep-ph]}}.

\bibitem{Fan:2026kxx}
J.~Fan and M.~Reece, ``{Higgsino Above the Sea of Fog},''
  \href{http://arxiv.org/abs/2609.01504}{{\ttfamily arXiv:2609.01504
  [hep-ph]}}.

\bibitem{Lou:2026idn}
Y.~Lou and C.-T. Lu, ``{Fermionic Dark Matter Absorption and the High-Energy
  Event in LUX-ZEPLIN},'' \href{http://arxiv.org/abs/2609.01592}{{\ttfamily
  arXiv:2609.01592 [hep-ph]}}.

\bibitem{Visinelli:2026kgt}
L.~Visinelli, ``{A Peccei-Quinn Origin for Inelastic Electroweak Dark Matter
  after LUX-ZEPLIN},'' \href{http://arxiv.org/abs/2609.02807}{{\ttfamily
  arXiv:2609.02807 [hep-ph]}}.

\bibitem{Yamashita:2026ump}
K.~Yamashita, ``{Inelastic Dark Photon Dark Matter for the LUX-ZEPLIN
  High-Recoil Event and the Galactic Halo Gamma-Ray Excess},''
  \href{http://arxiv.org/abs/2609.02868}{{\ttfamily arXiv:2609.02868
  [hep-ph]}}.

\bibitem{DiMauro:2026ldr}
M.~Di~Mauro, ``{Dark Matter at the Kinematic Edge: Interpreting the 248 keV LZ
  Nuclear-Recoil Candidate},''
  \href{http://arxiv.org/abs/2609.02608}{{\ttfamily arXiv:2609.02608
  [hep-ph]}}.

\bibitem{Nomura:2026qyq}
Y.~Nomura, ``{Dark Matter as the Z{\_}2 Partner of the Standard Model Higgs
  Boson},'' \href{http://arxiv.org/abs/2609.02505}{{\ttfamily arXiv:2609.02505
  [hep-ph]}}.

\bibitem{Smirnov:2026aqk}
J.~Smirnov, S.~Griffith, and J.~F. Beacom, ``{Inelastic Signatures of
  Electroweak Dark Matter},'' \href{http://arxiv.org/abs/2609.04144}{{\ttfamily
  arXiv:2609.04144 [hep-ph]}}.

\bibitem{Unwin:2026rdp}
J.~Unwin, ``{Axion Portal Dark Matter and the LUX-ZEPLIN High-Recoil Event},''
  \href{http://arxiv.org/abs/2609.04186}{{\ttfamily arXiv:2609.04186
  [hep-ph]}}.

\bibitem{McCabe:2026crm}
C.~McCabe, ``{Seasonal dark matter from the LUX-ZEPLIN high-energy event},''
  \href{http://arxiv.org/abs/2609.04181}{{\ttfamily arXiv:2609.04181
  [hep-ph]}}.

\bibitem{Jeesun:2026vzo}
S.~Jeesun and A.~Majumdar, ``{Atmospheric neutrino up-scattering explanation of
  LZ 2026 excess},'' \href{http://arxiv.org/abs/2609.04185}{{\ttfamily
  arXiv:2609.04185 [hep-ph]}}.

\bibitem{Rodd:2026tyn}
N.~L. Rodd, B.~R. Safdi, T.~R. Slatyer, and W.~L. Xu, ``{Confronting the
  Higgsino Interpretation of the LZ Event with the High-Energy Sideband},''
  \href{http://arxiv.org/abs/2609.04175}{{\ttfamily arXiv:2609.04175
  [hep-ph]}}.

\bibitem{Chattopadhyay:2026ryw}
U.~Chattopadhyay, D.~Das, R.~Puri, and J.~Roy, ``{Sub-TeV Singlino Dark Matter
  in light from Sagittarius A$^\ast$ and LUX-ZEPLIN Nuclear-Recoil Event},''
  \href{http://arxiv.org/abs/2609.02994}{{\ttfamily arXiv:2609.02994
  [hep-ph]}}.

\bibitem{Dent:2026bji}
J.~B. Dent and J.~L. Newstead, ``{Exothermic and Endothermic Inelastic Dark
  Matter Interpretations at LZ: Sideband Constraints and Future Prospects},''
  \href{http://arxiv.org/abs/2609.04673}{{\ttfamily arXiv:2609.04673
  [hep-ph]}}.

\bibitem{Gu:2026vto}
G.~Gu, L.~Li, S.-S. Tang, and Y.~Xu, ``{Inelastic from the Other Side: Xenon
  Excitation Signals in Light of the LZ High-Recoil Event},''
  \href{http://arxiv.org/abs/2609.05291}{{\ttfamily arXiv:2609.05291
  [hep-ph]}}.

\bibitem{deLima:2026shq}
C.~H. de~Lima, ``{Exothermic Dark Matter at LZ},''
  \href{http://arxiv.org/abs/2609.05204}{{\ttfamily arXiv:2609.05204
  [hep-ph]}}.

\bibitem{LHCSUSYXSWG}
{LHC SUSY Cross Section Working Group}, ``Nlo+nnll higgsino-like electroweakino
  production cross sections at 13 tev.''
  \url{https://twiki.cern.ch/twiki/bin/view/LHCPhysics/SUSYNLONNLLCrossSections13TeVHinoAll},
  2026.
\newblock Accessed September 2026.

\bibitem{Fukuda:2017jmk}
H.~Fukuda, N.~Nagata, H.~Otono, and S.~Shirai, ``{Higgsino Dark Matter or Not:
  Role of Disappearing Track Searches at the LHC and Future Colliders},''
  \href{http://dx.doi.org/10.1016/j.physletb.2018.03.088}{{\em Phys. Lett. B}
  {\bfseries 781} (2018) 306--311},
  \href{http://arxiv.org/abs/1703.09675}{{\ttfamily arXiv:1703.09675
  [hep-ph]}}.

\bibitem{ATLAS2026Disappearing}
{ATLAS Collaboration}, ``Search for long-lived charginos and tau-sleptons using
  final states with a disappearing track in pp collisions at sqrt(s)=13 tev
  with the atlas detector,''
  \href{http://dx.doi.org/10.1007/JHEP07(2026)152}{{\em JHEP} {\bfseries 07}
  (2026) 152}, \href{http://arxiv.org/abs/2603.08315}{{\ttfamily
  arXiv:2603.08315 [hep-ex]}}.

\bibitem{CMS2026SoftTrack}
{CMS Collaboration}, ``Search for electroweakinos in compressed-spectrum
  scenarios with low-momentum isolated tracks in proton-proton collisions at
  sqrt(s)=13 tev,'' \href{http://arxiv.org/abs/2604.25604}{{\ttfamily
  arXiv:2604.25604 [hep-ex]}}.

\bibitem{SaitoEtAl2019}
M.~Saito, R.~Sawada, K.~Terashi, and S.~Asai, ``Discovery reach for wino and
  higgsino dark matter with a disappearing track signature at a 100 tev pp
  collider,'' \href{http://dx.doi.org/10.1140/epjc/s10052-019-6974-2}{{\em Eur.
  Phys. J. C} {\bfseries 79} (2019) 469},
  \href{http://arxiv.org/abs/1901.02987}{{\ttfamily arXiv:1901.02987
  [hep-ph]}}.

\bibitem{CapdevillaEtAl2021}
R.~Capdevilla, F.~Meloni, R.~Simoniello, and J.~Zurita, ``Hunting wino and
  higgsino dark matter at the muon collider with disappearing tracks,''
  \href{http://dx.doi.org/10.1007/JHEP06(2021)133}{{\em JHEP} {\bfseries 06}
  (2021) 133}, \href{http://arxiv.org/abs/2102.11292}{{\ttfamily
  arXiv:2102.11292 [hep-ph]}}.

\end{thebibliography}\endgroup
\end{document}